\documentclass[reprint,
superscriptaddress,
showpacs, 
nofootinbib
amsmath,amssymb,
aps,
pra,
]{revtex4-2}
\usepackage{graphicx}
\usepackage{dcolumn}
\usepackage{bm}
\usepackage{dsfont}
\usepackage{physics}
\usepackage{bbold}
\usepackage{color}
\usepackage{soul}
\setstcolor{red}
\usepackage{ifthen}
\usepackage{enumitem}
\usepackage{url}
\usepackage[hidelinks]{hyperref}
\hypersetup{
    colorlinks,
    linkcolor={blue},
    citecolor={blue},
    urlcolor={blue}
}
\usepackage{url}
 \usepackage[thinc]{esdiff}
\newboolean{showannotations}   
\setboolean{showannotations}{false} 
\newcommand*{\added}[1]{%
  \ifthenelse{\boolean{showannotations}}{{\color{blue}#1}}{#1}%
}

\newcommand*{\removed}[1]{%
  \ifthenelse{\boolean{showannotations}}{\st{#1}}{}%
}

\newcommand*{\change}[2]{%
  \ifthenelse{\boolean{showannotations}}{{\color{red}\st{#1}}{\color{blue}#2}}{#2}%
}

\begin{document} 

\title{Microscopic Theory of Spin-Orbit Interactions of Light}
\author{Saeed Rahmanian Koshkaki}%
\affiliation{Department of Chemistry, Texas A\&M University, College Station, TX 77843, USA}
\author{Ding Xu}%
\affiliation{Department of Chemistry, Columbia University, New York, NY 10027, USA}
\author{Michael Fowler}%
\affiliation{Department of Chemistry, Texas A\&M University, College Station, TX 77843, USA}
\author{Milan Delor}
\affiliation{Department of Chemistry, Columbia University, New York, NY 10027, USA}
\author{Arkajit Mandal}%
\email{mandal@tamu.edu}
\affiliation{Department of Chemistry, Texas A\&M University, College Station, TX 77843, USA}

\begin{abstract}
\footnotesize
Optical spin-orbit interactions (SOI), demonstrated experimentally in optical cavities, enable the polarization degrees of freedom of light to be described and controlled as an effective pseudospin, providing a foundation for spintronic-like control of light and the development of next-generation optical technologies. However, prior works describe this emergent effect using phenomenological models that fail to uncover its microscopic quantum origins, and as a result, a rigorous connection between material electronic structure and the optical SOI remains elusive. In this work, we address this fundamental challenge by developing a microscopic quantum theory starting from the non-relativistic quantum electrodynamical (QED) description of light-matter interactions. We show that this effect arises from virtual interactions between confined photons and material excitations. Crucially, our theoretical framework provides microscopic material-design principles for tuning the distinct components of optical SOI, which include the optical birefringence, photonic self-energies, TE-TM splitting, and optical Zeeman splitting, enabling precise engineering of effective magnetic fields, diabolical points, and pseudospin textures. We anticipate that the predictive capability of our microscopic theory will enable systematic screening and design of material systems for photonic spintronics and optical technologies.
\end{abstract}

\maketitle
{\footnotesize

{\bf Introduction.} The creation and manipulation of spin degrees of freedom, traditionally electronic spins, are central to spintronics, offering low-power logic with minimal Joule heating, non‑volatile memory, and novel quantum functionalities.~\cite{jungwirth2012spin, vzutic2004spintronics, kang2015spintronics} Presently, realizing long-lived coherent spin currents, long-distance spin transport, and ultrafast tunability of spin polarization remain major challenges in spintronics.~\cite{liang2024polariton, gu2024challenges, guo2024emerging, han2014graphene} Synthetic spin-orbit coupling, generated through light-matter interactions, offers a unique opportunity to address these key challenges by enabling photonic spin-momentum locking and spin-based information encoding, while also benefiting from the ability of photons to maintain macroscopic coherence at room temperature.~\cite{liang2024polariton, xu2026giant, xiang2024optical, shi2025coherent, kapitanova2014photonic, high2015visible} This includes our recent work,~\cite{xu2026giant}  where we show that confined radiation within multilayered dielectric materials (layered van der Waals materials) can generate such synthetic spin-orbit coupling of light, leading to spin-polarized transport of photons. This effect arises because different polarizations of light acquire distinct group velocities, as shown in Fig.~\ref{Fig1_schemband}b-c. In Fig.\ref{Fig1_schemband}b-c, the right- and left-hand polarized (RHP and LHP) modes exhibit polarization-dependent dispersions, such that excitation near the diabolical point (where two dispersions cross) accesses branches with opposite slopes and hence opposite group velocities.~\cite{liang2024polariton, Kavokin_OSHE_2005, xu2026giant} The resulting separation of the two polarized components leads to spin-polarized photon transport, as illustrated in Fig.~\ref{Fig1_schemband}a. Overall, this optical spin Hall effect (OSHE) is emerging as a promising platform for spin-based applications including chip-compatible polarizers and nonlinear elements for optical computing.

\begin{figure} 
\centering
\includegraphics[width=1\linewidth]{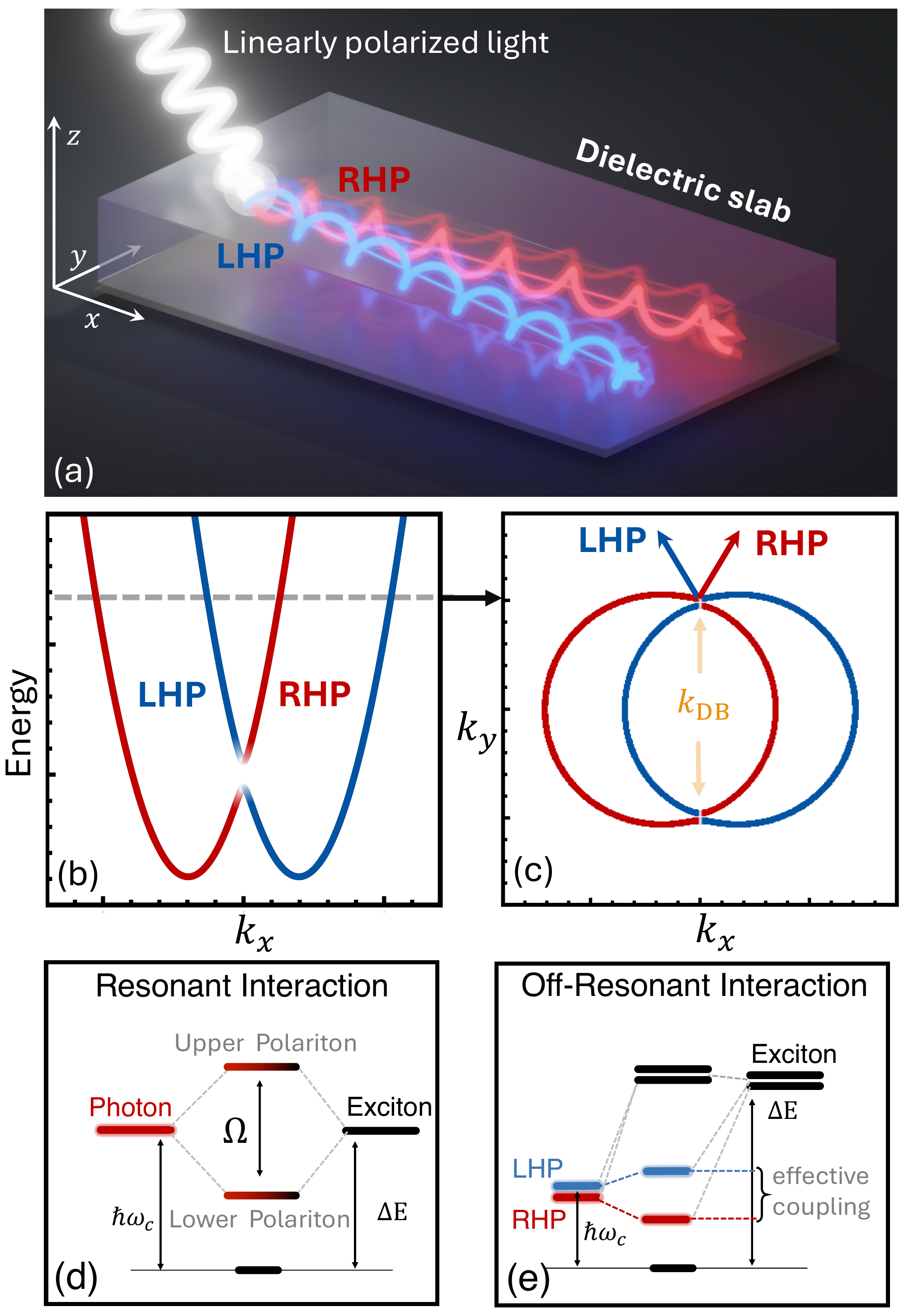}
\caption{\footnotesize \textbf{Schematic Illustration of Light Polarization Splitting and Low-Energy Photonic Dispersion.} (a) Illustration of RHP and LHP polarization mode splitting inside a thin slab of anisotropic material. (b) Energy dispersion cut near the diabolical-like point $k_{\text{DB}}$. (c) Energy dispersion at a fixed energy with $k_{\text{DB}}$ represented in yellow. Arrows indicate the propagating wavevector direction (dispersion slope) of each polarization mode. (d) Exciton-polariton formation when the exciton and photon are in resonance. (e) Exciton at high energy uses virtual interactions to split RHP and LHP photonic modes. }
\label{Fig1_schemband}
\end{figure}

However, the microscopic origin of optical spin-orbit interactions remains unclear. Prior works describe the observed photonic spin-orbit interactions through an analogy with the electronic spins and their spin-orbit couplings using the following phenomenological $2 \times 2$  Rashba-Dresselhaus model expressed in the circularly polarized basis~\cite{liang2024polariton, xiang2024optical, spencer2021spin, bleu2018measuring}: 

\begin{equation}\label{RD}
\hat{H}_\mathrm{RD} = 
\begin{pmatrix}
\omega_{\boldsymbol{k}} +\delta_z &\alpha+\beta (k_x+ik_y)^2  \\
 \alpha+\beta (k_x-ik_y)^2& \omega_{\boldsymbol{k}} - \delta_z
\end{pmatrix},
\end{equation}

where $\omega_{\boldsymbol{k}}$ is the photon frequency associated with the in-plane wavevector ${\boldsymbol{k}} = k_x\vec{\boldsymbol{e}}_x+k_y\vec{\boldsymbol{e}}_y = |{\boldsymbol{k}}| \vec{\boldsymbol{e}}_\parallel$, $\alpha$ is referred to as the optical birefringence, $\beta$ is the TE-TM splitting, and $\delta_z$ is the Zeeman splitting, all of which characterize photonic spin-orbit interactions \cite{bleu2018measuring, Terifmmode_GaugeField_2014, spencer2021spin, liang2024polariton}. These parameters are typically extracted by fitting the experimentally measured dispersion curves; therefore, their connection to material electronic structure remains unknown. The elucidation of this relationship is essential to systematically engineer and tune photonic spin-orbit interactions via material design. 

In this work, we resolve this fundamental challenge by developing a microscopic quantum mechanical theory starting from the non-relativistic QED description of an extended material coupling to confined radiation, formulated in the dipole gauge beyond the long-wavelength approximation. We show that the spin-orbit couplings of photons emerge from virtual interactions to material excitonic states, analogous to a superexchange mechanism~\cite{berkelbach2013microscopic, anderson1959new}, and that their strength and symmetry are dictated directly by the energies of the underlying excitonic states, their transition dipole moments, and their spatial orientations. Overall, our microscopic theory provides a direct analytical expression connecting material properties with optical birefringence, TE-TM splitting, and Zeeman splitting. Using our theory, we predict the location of the diabolical points in various scenarios and compute the effective magnetic field interacting with the photonic pseudospins. We present our results for four distinct materials and show how different regimes of optical spin-orbit couplings emerge from the respective microscopic electronic structures, establishing material-level design principles for controlling photonic spin transport. We also emphasize that, while a classical treatment that uses dielectric tensors can reproduce the spectral features of OSHE, it does not provide the microscopic material design principles provided in the present work. Moreover, the present quantum mechanical framework can be naturally extended to describe photonic spin coherence and its phonon-induced decoherence under ambient conditions.



 
{\bf Theory.} 
To uncover the microscopic origin of the spin-orbit interactions of light, we consider a prototypical material featuring multiple excitonic states coupling to confined radiation modes, as realized either in a fabricated Fabry–P\'{e}rot cavity or intrinsically within the dielectric material itself via self-hybridization. We consider a cavity quantization length $L$ along the $z$-direction, which corresponds to the cavity mirror spacing or the thickness of dielectric material with the cavity setup extending in the $xy$-plane. The solutions of the Maxwell equations for this system reveal two different polarization modes: transverse electric (TE or $s$) and transverse magnetic (TM or $p$), with each mode associated to an in-plane wavevector of ${\boldsymbol{k}} = k_x\vec{\boldsymbol{e}}_x+k_y\vec{\boldsymbol{e}}_y$. The light-matter interaction of TE and TM modes in such a system can be modeled using a generalized Tavis–Cummings Hamiltonian, where exciton-photon couplings are considered beyond the long-wave approximation, written as~\cite{Krupp2025FullQ,rahmanian2026Multilayer, xu2026giant, KeelingARPC2020, SunJPCC2025, manjalingal2025tilted, MandalCR2023, sarkar2026non}
\begin{equation}
\hat{H}_{\vec{\boldsymbol{\mu}} \cdot \vec{\boldsymbol{E}}} = \hat{H}_{\text{Ph}}+\hat{H}_{\text{M}}+\hat{H}_{\text{I}}\,.
\end{equation}
Here, $\hat{H}_{\text{Ph}}$, $\hat{H}_{\text{M}}$, and $\hat{H}_{\text{I}}$ represent the photon, exciton, and exciton-photon interaction terms. The photonic Hamiltonian~\cite{rahmanian2026Multilayer, SunJPCC2025, LitinskayaPSS2004, michetti2005polariton} (using atomic units, with $\hbar=1$) in the circularly polarized basis is given by  

\begin{equation}
\hat{H}_{\text{Ph}} = \sum_{\boldsymbol{k}} \omega_{\boldsymbol{k}}\left(\hat{a}^\dagger_{\boldsymbol{k}}\hat{a}_{\boldsymbol{k}} + \hat{b}^\dagger_{\boldsymbol{k}}\hat{b}_{\boldsymbol{k}}  \right)\, ,
\end{equation}
where $\hat{a}^\dagger_{\boldsymbol{k}}$ ($\hat{b}^\dagger_{\boldsymbol{k}}$) are the creation operators for the RHP (LHP) photon modes, with corresponding photon energy (in the case of perfectly reflective mirrors in a Fabry-P\'{e}rot cavity) of $\omega_{\boldsymbol{k}}=\sqrt{{c^2}|\boldsymbol{k}|^2+\omega_0^2}$ where $c = c_0/\eta$ is the speed of light inside a dielectric medium with a refractive index of $\eta$. Note that here we consider $\omega_{\boldsymbol{k}}$ to be polarization-independent, which is strictly true for an ideal planar optical cavity bounded by perfectly conducting mirrors. In realistic optical microcavities, such as distributed Bragg reflectors (DBRs),~\cite{panzarini1999exciton, Kavokin_OSHE_2005, Rupprecht2DTETMsplitting2021} this degeneracy is broken due to polarization-dependent optical confinement, which can be viewed as an effective spin-orbit interaction. However, this effect originates from the boundary conditions of the Helmholtz equation and relates to the tails of the confined radiation. Consequently, the magnitude of such an effect is negligible compared to the light-matter interaction induced optical SOI discussed here. Furthermore, optical SOI generated by light-matter interaction is a technologically promising platform as the effect is tunable via material design. Here, $\omega_0 = \pi c/ L$. Note that here RHP and LHP are linear combinations of TE and TM polarization, following the convention used in previous work \cite{bleu2018measuring, gianfrate2020ExpQG, Hu2024QGT}. For the matter part, we use a tight-binding model describing a rectangular lattice, with $N_x$ sites along the $x$-direction and $N_y$ sites along the $y$-direction. Assuming periodic boundary conditions in the $xy$-plane, we can define the in-plane wavevectors of $k_x = \frac{2\pi n_x}{N_xa_x}$ for $n_x \in \{0,1,\dots, N_x-1\}$ and $k_y = \frac{2\pi n_y}{N_ya_y}$ for $n_y \in \{0,1,\dots, N_y-1\}$, where $a_x$ and $a_y$ are lattice constants along the $x$ and $y$ directions, respectively. Assuming $N_d$ excitonic states per unit cell, the matter Hamiltonian in  the reciprocal space is given by

\begin{equation}
H_{\text{M}} = \sum_{d}^{N_d}\left(\sum_{\boldsymbol{k}}\epsilon_{\boldsymbol{k},d}X^\dagger_{\boldsymbol{k},d}X_{\boldsymbol{k},d}\right) \approx \sum_{d}^{N_d}\left(\sum_{\boldsymbol{k}}{\epsilon}_{d}X^\dagger_{\boldsymbol{k},d}X_{\boldsymbol{k},d}\right) \,,
\end{equation}
where $X^\dagger_{\boldsymbol{k},d}$ creates $d$'th exciton of wavevector $\boldsymbol{k}$ with an energy of $\epsilon_{\boldsymbol{k},d}$. Since the photonic dispersion is much steeper than the typical excitonic dispersion, we approximate the latter as dispersionless, $\epsilon_{\boldsymbol{k},d} \approx {\epsilon}_{d}$. The interaction between the excitons and photons, after applying a rotating-wave approximation in the circularly polarized basis, is given by \cite{rahmanian2026Multilayer, SunJPCC2025, LitinskayaPSS2004, MandalNL2023}
\begin{equation}\label{OriginalModelSPseparate}
\begin{split}
\hat{H}_{\text{I}} 
=&\sum_{\boldsymbol{k},d}\left(g_{d}^{+}(\boldsymbol{k})\hat{X}_{\boldsymbol{k},d}\hat{a}_{\boldsymbol{k}}^\dagger + g_{d}^{-}(\boldsymbol{k})\hat{X}_{\boldsymbol{k},d}\hat{b}_{\boldsymbol{k}}^\dagger\right)+h.c.\\
\end{split}
\end{equation}
where $g_{d}^{+}(\boldsymbol{k})$  and $g_{d}^{-}(\boldsymbol{k})$ characterize the light-matter coupling strength between the $d$th exciton to RHP and LHP photon modes, respectively. Considering perfectly reflective mirrors (although our results are not limited to this assumption), these coupling terms are written as

\begin{align}
g_{d}^{\pm}(\boldsymbol{k})=& -\frac{g_0}{\sqrt{2}|\boldsymbol{k}|}\Bigg(\sqrt{\frac{\omega_{\boldsymbol{k}}}{\omega_0}}\left( \mu^{d}_x{k_y}  -\mu^{d}_y k_x \right) \pm i\sqrt{\frac{\omega_0}{\omega_{\boldsymbol{k}}}} \left( \mu^{d}_x{k_x} + \mu^{d}_y k_y \right)\Bigg) \nonumber\\
&\mp  g_0\frac{c|\boldsymbol{k}|\mu^{d}_{z}}{\sqrt{2\omega_0\omega_{\boldsymbol{k}}}},
\end{align}
where $\vec{\boldsymbol{\mu}}_d = \mu_x^d\vec{\boldsymbol{e}}_x+\mu_y^d\vec{\boldsymbol{e}}_y+\mu_z^d\vec{\boldsymbol{e}}_z = \vec{\boldsymbol{\mu}}^d_\parallel+\mu_z^d\vec{\boldsymbol{e}}_z$ is the transition dipole moment of the $d$th exciton, and $g_0$ is the overall light-matter coupling constant. Further details are provided in the SI.

{\bf Effective Hamiltonian.} 
To microscopically derive an effective photonic Hamiltonian that describes photonic spin-orbit interactions, we start with the non-relativistic QED light-matter Hamiltonian presented in the preceding section and perform a Schrieffer-Wolff transformation to remove the excitonic degrees of freedom. In doing so, we introduce effective coupling terms between the photonic polarizations. We find that this effective photonic spin-orbit interaction does not take the form of the phenomenological Rashba-Dresselhaus Hamiltonian, shown in Eq.~\ref{RD} and adopted in recent works~\cite{liang2024polariton, xiang2024optical, spencer2021spin, bleu2018measuring, gianfrate2020ExpQG}, but instead assumes a generalized Rashba-Dresselhaus (GRD) form in which the birefringence, TE-TM splitting, and Zeeman-like interaction parameters acquire a momentum dependence. Specifically, the generalized Rashba-Dresselhaus Hamiltonian introduced in this work and obtained via the Schrieffer-Wolff transformation \cite{schriefferWolff1966original, Bravyi2011SW} is computed up to second order and written as 
\begin{equation}\label{GRD}
\hat{\mathcal{H}}_\mathrm{GRD} = \mathcal{P} e^{\hat{S}}\hat{H}e^{-\hat{S}} \mathcal{P} \approx \mathcal{P}\hat{H}_{0}\mathcal{P}+\frac{1}{2}\mathcal{P}\left[\hat{S},\hat{H}_{\text{I}}\right]\mathcal{P},
\end{equation}
where $\hat{H}_{0} = \hat{H}_{\text{Ph}}+\hat{H}_{\text{M}}$, $\hat{S}$ is a skew-Hermitian operator,  and $\mathcal{P} = \sum_{\boldsymbol{k}} \hat{a}^\dagger_{\boldsymbol{k}}|0\rangle\langle0| \hat{a}_{\boldsymbol{k}} + \hat{b}^\dagger_{\boldsymbol{k}}|0\rangle\langle0| \hat{b}_{\boldsymbol{k}}$ is a projection operator that projects into the photonic subspace (here $\ket{0}$ is the vacuum state). We find that the effective spin-orbit-coupled photonic Hamiltonian can be expressed as a generalized Rashba-Dresselhaus Hamiltonian of the form
\begin{align}
\hat{\mathcal{H}}_\mathrm{GRD} =  \begin{pmatrix}
{\omega}_{\boldsymbol{k}} + \Sigma({\boldsymbol{k}}) +\delta_z({\boldsymbol{k}})& \alpha (\boldsymbol{k}) + \beta (\boldsymbol{k}) \left(k_x+ik_y\right)^2  \\
\alpha (\boldsymbol{k}) + \beta^* (\boldsymbol{k}) \left(k_x-ik_y\right)^2 & {\omega}_{\boldsymbol{k}} + \Sigma({\boldsymbol{k}}) - \delta_z({\boldsymbol{k}})
\end{pmatrix}
\end{align}
where ${\omega}_{\boldsymbol{k}} + \Sigma({\boldsymbol{k}})$ is the renormalized photonic frequency, $\alpha(\boldsymbol{k})$ is the generalized  optical birefringence, $\beta(\boldsymbol{k})$ is the generalized TE-TM splitting, and $\delta_z(\boldsymbol{k})$ is the generalized Zeeman splitting, all of which now acquire in-plane momentum dependence (with their explicit expressions provided below). To elucidate the different contributions to the photonic spin-orbit interaction, we decompose the effective Hamiltonian according to perturbative order as 

\begin{equation}\label{FullSOIH}
\begin{split}
\hat{\mathcal{H}}_{\text{GRD}} = {\omega}_{\boldsymbol{k}} \mathds{1} + \hat{\mathcal{H}}_0^{\text{SOI}} + \hat{\mathcal{H}}_1^{\text{SOI}}
\end{split}
\end{equation}
where $\mathds{1}$ is the $2\times2$ identity matrix, and $\hat{\mathcal{H}}_0^{\mathrm{SOI}}$ and $\hat{\mathcal{H}}_1^{\mathrm{SOI}}$ denote the leading ($\propto (|\boldsymbol{k}|/\omega_0)^n$ for $n = 0$) and higher-order ($\propto (|\boldsymbol{k}|/\omega_0)^n$ for $n\ge 1$) photonic interaction terms, respectively. The leading photonic spin-orbit interaction term $\hat{\mathcal{H}}_0^{\text{SOI}}$ for $n=0$ has the form 
\begin{equation}
\begin{split}
\hat{\mathcal{H}}_0^{\text{SOI}} = 
\begin{pmatrix}
{\Sigma}_0& \beta_0(\boldsymbol{k})\left(k_x+ik_y\right)^2 \\
\beta^*_0({\boldsymbol{k}})\left(k_x-ik_y\right)^2 & {\Sigma}_0
\end{pmatrix}\, .
\end{split}
\end{equation}
Here, ${\Sigma}_0=g_0^2\sum_d  {|\vec{\boldsymbol{\mu}}_\parallel^d|^2}/{2\Delta_d}$ is the leading spin-independent self energy, $\beta_0(\boldsymbol{k})=-\frac{g_0^2}{|\boldsymbol{k}|^2}\sum_d {\left(\vec{\boldsymbol{\mu}}^d_\parallel \cdot \vec{\boldsymbol{e}}_- \right)^2}/{\Delta_d}$, for $\Delta_d = \omega_0 - {\epsilon}_d$ and $\vec{\boldsymbol{e}}_- =  (\vec{\boldsymbol{e}}_x-i\vec{\boldsymbol{e}}_y)/\sqrt{2}$. The higher order term, represented by $\hat{\mathcal{H}}_1^{\text{SOI}}$, is written as 
\begin{equation}
\begin{split}
\hat{\mathcal{H}}_1^{\text{SOI}} = 
\begin{pmatrix}
{\Sigma}_1({\boldsymbol{k}})+\delta_z(\boldsymbol{k})& \alpha(\boldsymbol{k}) + \beta_1\left(k_x+ik_y\right)^2 \\
\alpha(\boldsymbol{k}) + \beta^*_1\left(k_x-ik_y\right)^2 & {\Sigma}_1(\boldsymbol{k})-\delta_z(\boldsymbol{k})
\end{pmatrix}\, ,
\end{split}
\end{equation}
where the explicit form of the higher order spin-independent self-energy $ \Sigma_1({\boldsymbol{k}})$ is written as
\begin{widetext}
\begin{align}
\Sigma_1({\boldsymbol{k}}) =& \frac{g_0^2}{2}\sum_d \left( \Bigg[ \frac{c^2|{\boldsymbol{k}}|^2}{2\omega_0^2}\left( \left(  (\vec{\boldsymbol{\mu}}^d_\parallel\times \vec{\boldsymbol{e}}_\parallel)\cdot\vec{\boldsymbol{e}}_z\right)^2-\left( \vec{\boldsymbol{\mu}}^d_\parallel\cdot \vec{\boldsymbol{e}}_\parallel \right)^2 \right) + (\mu^d_{z})^2 \left(\frac{c^2|{\boldsymbol{k}}|^2}{\omega_0^2} - \frac{c^4|{\boldsymbol{k}}|^4}{2{\omega_0^4}}  \right) \Bigg]\left( \frac{1}{\Delta_d} - \frac{c^2|{\boldsymbol{k}}|^2}{2\omega_0\Delta_d^2}  \right) - \frac{|\vec{\boldsymbol{\mu}}^d_\parallel|^2c^2|{\boldsymbol{k}}|^2}{2\omega_0\Delta_d^2} \right)\, ,
\end{align}
\end{widetext}
such that the renormalized photonic frequency  is written as ${\omega}_{\boldsymbol{k}} + \Sigma({\boldsymbol{k}}) =  {\omega}_{\boldsymbol{k}} + \Sigma_0 + \Sigma_1({\boldsymbol{k}})$. The generalized optical birefringence term is written as 
\begin{align}
    \alpha({\boldsymbol{k}}) =& \frac{g_0^2}{2}\sum_d \left( \frac{1}{\Delta_d} - \frac{c^2|{\boldsymbol{k}}|^2}{2\omega_0\Delta_d^2}  \right) \nonumber\\
    &\times\left( |\vec{\boldsymbol{\mu}}^d_\parallel|^2\frac{c^2 |{\boldsymbol{k}}|^2}{2\omega_0^2} - (\mu^d_{z})^2 \left(\frac{c^2|{\boldsymbol{k}}|^2}{{\omega_0^2}} - \frac{c^4|{\boldsymbol{k}}|^4}{2{\omega_0^4}}  \right) \right).
\end{align}
Interestingly, this expression reveals that, through careful engineering of the material transition dipoles, for example, by designing a custom multilayered heterostructure~\cite{geim2013VanderWaals, castellanos2022VanderWaals, sortino2025vanderWaalsOpticC, luo2024vDWStrongCoupling, liang2024polariton}, the optical birefringence term could be selectively suppressed when $\sum_d |\vec{\boldsymbol{\mu}}^d_\parallel|^2 \approx 2\sum_d (\mu_z^d)^2$ (assuming equal/similar $\Delta_d$ for all excitonic states and $|{\boldsymbol k}| \rightarrow 0$). The generalized (optical) Zeeman-splitting term is
\begin{align}
\delta_z({\boldsymbol{k}}) =&  \frac{cg_0^2}{{\omega_0}}\sum_d  {\left(  (\vec{\boldsymbol{\mu}}^d_\parallel\times \boldsymbol{k})\cdot\vec{\boldsymbol{\mu}}_d\right)}    \left( \frac{1}{\Delta_d} - \frac{c^2|{\boldsymbol{k}}|^2}{2\omega_0\Delta_d^2}  \right),
\end{align}
which is nonzero only if the material possesses out-of-plane transition dipoles. Finally, the higher-order TE-TM splitting term is
\begin{align}
    \beta_1 =& \frac{g_0^2}{2}\sum_d \frac{c^2}{\omega_0\Delta_d^2} \left( \vec{\boldsymbol{\mu}}^d_\parallel \cdot \vec{\boldsymbol{e}}_- \right)^2,
\end{align}
such that the generalized TE-TM splitting term is written as $ \beta(\boldsymbol{k}) = \beta_0(\boldsymbol{k}) + \beta_1$. By tuning these parameters via material architecture, a specific SOI regime can be achieved to selectively control photonic spin transport.

Using the pseudospin picture of RHP and LHP, $\hat{\mathcal{H}}_\mathrm{GRD}$ can be interpreted as a photonic pseudospin interacting with an effective magnetic field, i.e.,

\begin{equation}
\hat{\mathcal{H}}_\mathrm{GRD} = (\omega_{\boldsymbol{k}}+\Sigma(\boldsymbol{k})) \mathds{1} + \vec{\boldsymbol{\sigma}}\cdot \boldsymbol{G}(\boldsymbol{k})\,,
\end{equation}
where $\vec{\boldsymbol{\sigma}}=(\sigma_x, \sigma_y, \sigma_z)$ are Pauli matrices in the circularly polarized photonic basis, and $\vec{\boldsymbol{G}}(\boldsymbol{k})=(G_x(\boldsymbol{k}), G_y(\boldsymbol{k}), G_z(\boldsymbol{k}))$ is the effective magnetic field with each of its component expressed as

\begin{equation}
\begin{split}
G_x(\boldsymbol{k}) &= \alpha(\boldsymbol{k}) + \operatorname{Re}[\beta(\boldsymbol{k})](k_x^2-k_y^2) - 2\operatorname{Im}[\beta(\boldsymbol{k})]k_xk_y\\
G_y(\boldsymbol{k}) &= \operatorname{Im}[\beta(\boldsymbol{k})](k_x^2-k_y^2) + 2\operatorname{Re}[\beta(\boldsymbol{k})]k_xk_y\\
G_z(\boldsymbol{k}) &= \delta_z(\boldsymbol{k})\,.
\end{split}
\end{equation}

We also define $\vec{\boldsymbol{G}}_\parallel(\boldsymbol{k}) = G_x(\boldsymbol{k})\vec{\boldsymbol{e}}_x + G_y(\boldsymbol{k})\vec{\boldsymbol{e}}_y$, which is the in-plane component of the magnetic field and $G_z(\boldsymbol{k})$ is the out-of-plane component. 

{\bf Diabolical Point.} The spin-polarized dispersion branches cross at the diabolical point, near which the spatial spin-polarization becomes maximal. In addition to this, while beyond the scope of the present work, diabolical points open up the possibilities for exploring exotic topological phenomena, including nontrivial Berry phase and Berry-curvature-driven anomalous Hall effects, which have been observed experimentally.~\cite{Bleu2018QG, gianfrate2020ExpQG, mavrotsoupakis2025topological, su2021direct} Our theory provides the location of the diabolical points, which can be found by solving the following equations 
\begin{subequations}\label{eq:diabolical}
\begin{align}
\left(\frac{u^2}{v}+4v\right)k_xk_y
-w|\boldsymbol{k}|^4-\bar{w}|\boldsymbol{k}|^6
&=0, \qquad v\neq 0,
\label{eq:diabolical_a}\\[6pt]
u(k_y^2-k_x^2)+w|\boldsymbol{k}|^4
+\bar{w}|\boldsymbol{k}|^6
&=0, \qquad v=0.
\label{eq:diabolical_b}
\end{align}
\end{subequations}
with the additional condition $\xi_xk_y = \xi_yk_x$. Here, $u=\sum_d ((\mu^d_x)^2 -(\mu_y^d)^2)/\Delta_d $, $v= \sum_d \mu^d_x\mu_y^d/\Delta_d$, $w=\sum_d c^2\left(|\boldsymbol{\mu}^d_\parallel|^2 - 2(\mu_d^z)^2\right)/(2\omega_0^2\Delta_d)$, $\Bar{w}=\sum_d c^4 (\mu_d^z)^2/(2\omega_0^2\Delta_d)$, and $\xi_{x/y} = \sum_d \mu^d_{z}  \mu^d_{x/y}/\Delta_d$ are microscopic material parameters that directly dictate the location of the diabolical point.  It is important to note that the conditions derived here are based on the {\it perturbative} effective Hamiltonian presented in Eq.~\ref{GRD} and, therefore, do not necessarily establish the existence of {\it exact} degeneracies in the full light-matter Hamiltonian as higher-order terms may lift this degeneracy to create a (diabolical) avoided crossing. Such avoided crossings, just as the diabolical points, are technologically important as excitation near these locations can enable pronounced polarization-dependent propagation.

\begin{figure*}[t]
\centering
\includegraphics[width=1.0\linewidth]{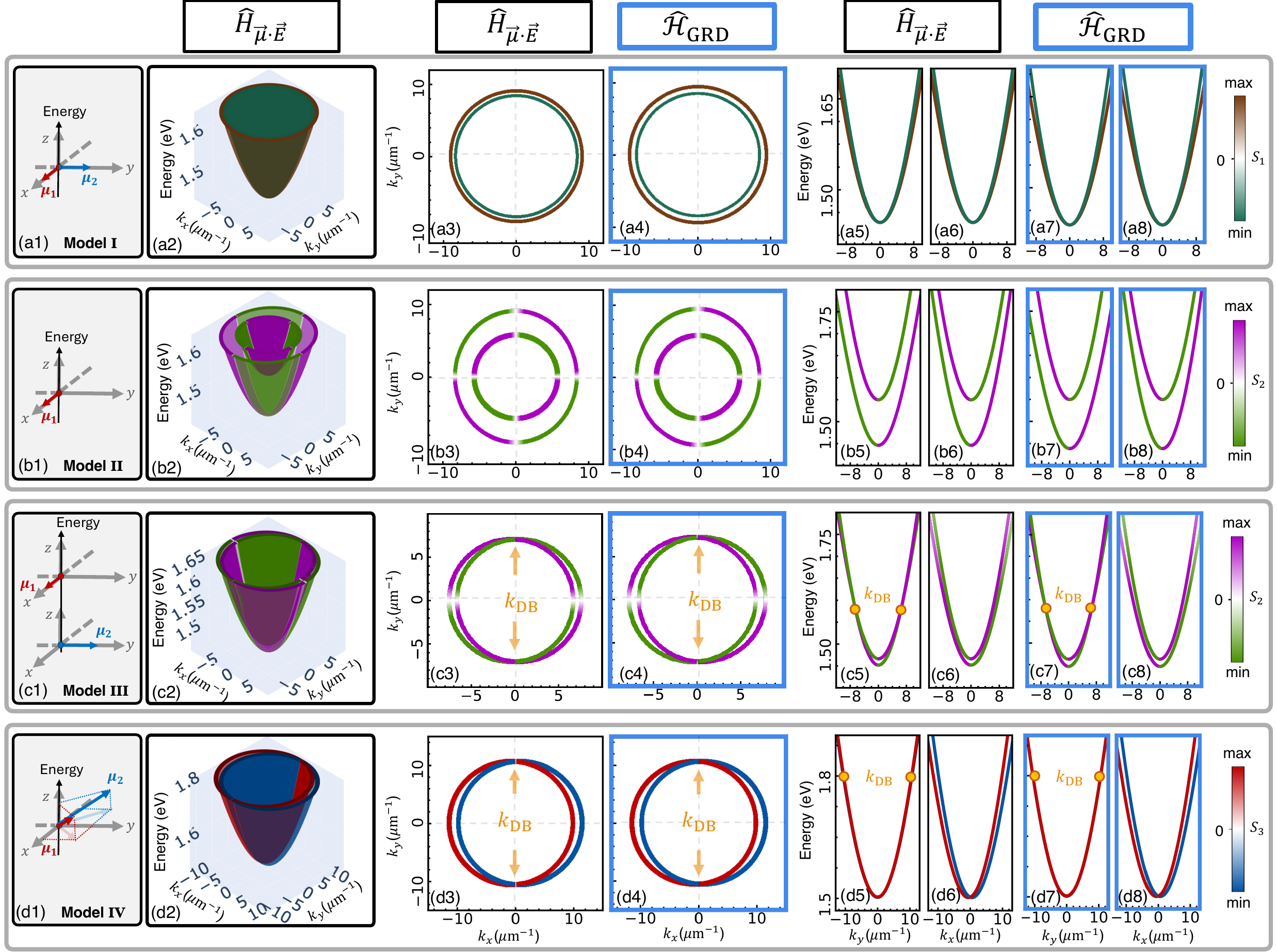}
\caption{\textbf{Energy Dispersion of $\hat{\mathcal{H}}_{\text{GRD}}$ Versus $\hat{H}_{\vec{\boldsymbol{\mu}} \cdot \vec{\boldsymbol{E}}}$ for Different Model Materials.} (a1) Schematic illustration of transition dipole moment orientation for Model I. (a2) Isometric view of energy dispersion of $\hat{H}_{\vec{\boldsymbol{\mu}} \cdot \vec{\boldsymbol{E}}}$ for Model I. (a3-4) cut of energy dispersion of $\hat{H}_{\vec{\boldsymbol{\mu}} \cdot \vec{\boldsymbol{E}}}$ and $\hat{\mathcal{H}}_{\text{GRD}}$  at $E_0 = 1.65$ eV, respectively. (a5-6) Cut of energy dispersion of $\hat{H}_{\vec{\boldsymbol{\mu}} \cdot \vec{\boldsymbol{E}}}$ at $k_x = 0.1 ~\mu\text{m}^{-1}$ and $k_y = 0.1 ~\mu\text{m}^{-1}$, respectively. (a7-8) Cut of energy dispersion of $\hat{\mathcal{H}}_{\text{GRD}}$ at $k_x = 0.1 ~\mu\text{m}^{-1}$ and $k_y = 0.1 ~\mu\text{m}^{-1}$, respectively. (b1-8) The same illustrations as (a1-8) but using the dipole orientation of Model II. (c1-8) The same illustrations as (a1-8) but using the dipole orientations of Model III and an energy cut at $E_0 = 1.6$ eV, with ${k}_y^{\text{DB}}\approx\pm 7.21 ~\mu\text{m}^{-1}$. (d1-8) The same illustrations as (a1-8) but using the dipole orientations of Model IV and energy cut at $E_0 = 1.8$ eV, with ${k}_y^{\text{DB}}\approx\pm 10.55 ~\mu\text{m}^{-1}$.}
\label{fig2_Alldispersion}
\end{figure*}
{\bf Results.} To elucidate how microscopic material parameters control the photonic interaction terms $\Sigma({\boldsymbol{k}}), \alpha({\boldsymbol{k}})$, $\beta({\boldsymbol{k}})$ and $\delta_z({\boldsymbol{k}})$, we consider four distinct material systems. For all cases, we assume the material is a dielectric slab that extends in the $xy$ plane and has a thickness $L=200$ nm along the $z$ direction, with lattice constants $a_x=a_y=5\,\mathrm{\AA}$.We assume a refractive index of $\eta=2$ and a light-matter coupling constant of $g_0=0.007349$ a.u., corresponding to a light-matter coupling of  $g_0 = 0.2$ eV for a transition dipole moment of $1$ a.u. Specifically, we consider the following four model material systems:
\begin{itemize}[label=, leftmargin=0.2cm]
    \item {\bf Model I:} An {\it isotropic} material consisting of two excitonic dipoles per unit cell with equal on-site energies of $\epsilon_1 = \epsilon_2=3$ eV and transition dipole moments of $\vec{\boldsymbol{\mu}}_1=2\vec{\boldsymbol{e}}_x$ and $\vec{\boldsymbol{\mu}}_2=2\vec{\boldsymbol{e}}_y$, schematically represented in Fig. \ref{fig2_Alldispersion}a1.
    \item {\bf Model II:} A {\it uniaxial} (nematic-like) material consisting of one excitonic dipole per unit cell with an on-site energy of $\epsilon_1 = 3$ eV and a transition dipole moment of $\vec{\boldsymbol{\mu}}_1=2\vec{\boldsymbol{e}}_x$, schematically represented in Fig. \ref{fig2_Alldispersion}b1.
    \item {\bf Model III:} An {\it anisotropic} material consisting of two excitonic dipoles per unit cell with on-site energies of $\epsilon_1 = 3.4$ eV and $\epsilon_2=3.1$ eV and transition dipole moments of $\vec{\boldsymbol{\mu}}_1=2\vec{\boldsymbol{e}}_x$ and $\vec{\boldsymbol{\mu}}_2=2\vec{\boldsymbol{e}}_y$, schematically represented in Fig. \ref{fig2_Alldispersion}c1.
    \item {\bf Model IV:} A {\it canted} material consisting of two excitonic dipoles per unit cell with equal on-site energies of $\epsilon_1 = \epsilon_2=3.6$ eV and transition dipole moments of $\vec{\boldsymbol{\mu}}_1=1.1\vec{\boldsymbol{e}}_x+1.1\vec{\boldsymbol{e}}_y+1.25\vec{\boldsymbol{e}}_z$ and $\vec{\boldsymbol{\mu}}_2=-1.1\vec{\boldsymbol{e}}_x+1.1\vec{\boldsymbol{e}}_y+1.25\vec{\boldsymbol{e}}_z$, schematically represented in Fig. \ref{fig2_Alldispersion}d1.  
\end{itemize}

In Fig.~\ref{fig2_Alldispersion}, we present the spin-polarized photonic dispersions and compare the predictions of the generalized Rashba-Dresselhaus ($\hat{\mathcal{H}}_\mathrm{GRD}$) model against the non-relativistic QED description via the dipole gauge Hamiltonian $\hat{H}_{\vec{\boldsymbol{\mu}}\cdot\vec{\boldsymbol{E}}}$. Here, we compute the Stokes vector (see details in the SI), $\vec{S} = (S_0, S_1, S_2, S_3)$, to characterize the photonic polarization and spin-momentum locking of the dispersions.

\begin{figure*}[ht]
    \centering
    \includegraphics[width=1\linewidth]{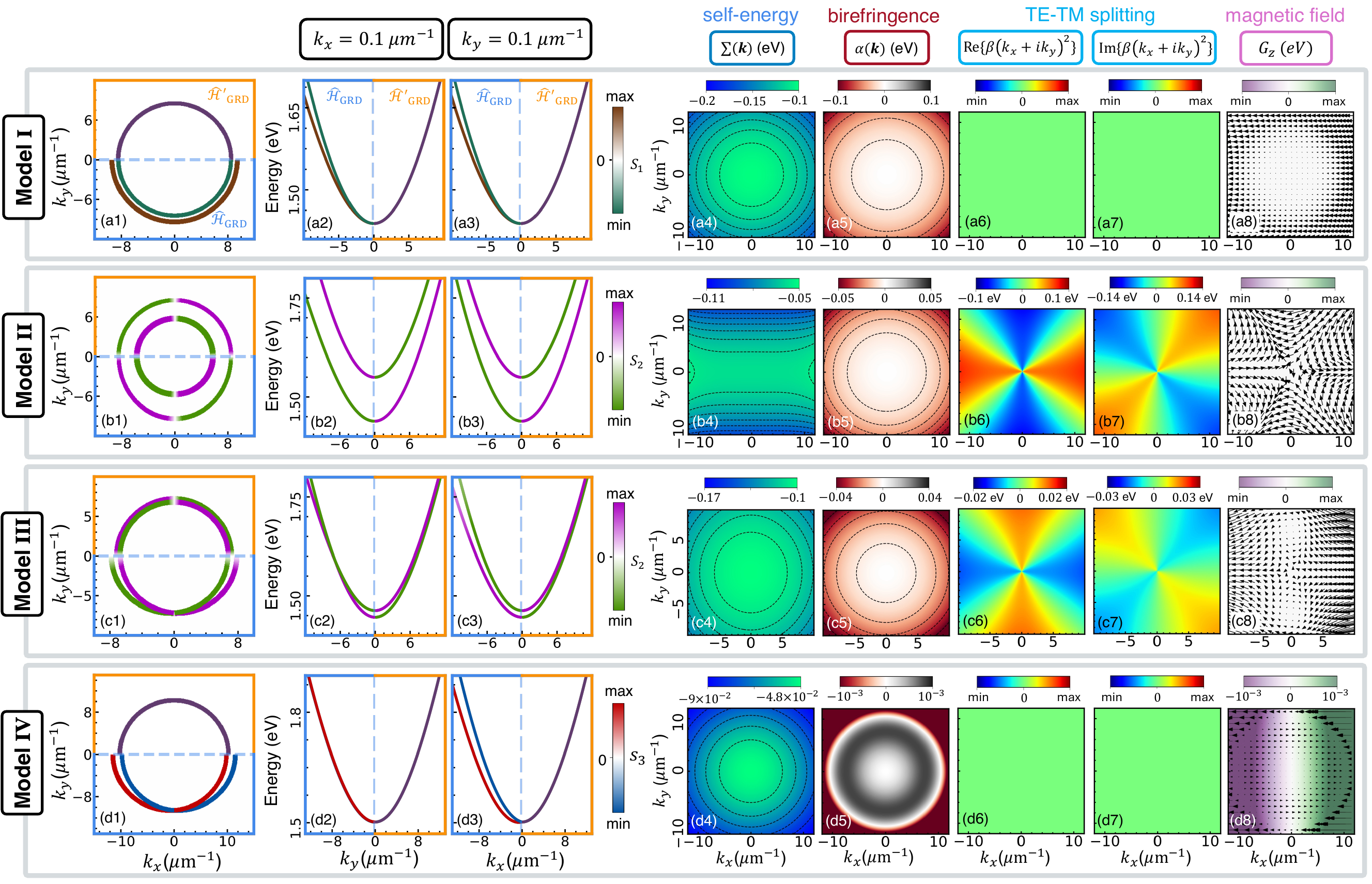}
    \caption{\textbf{Parameters of $\hat{\mathcal{H}}_\mathrm{GRD}$ for Different Model Materials.} (a1) Energy dispersion of $\hat{\mathcal{H}}_\mathrm{GRD}$ for Model I with the $\hat{\mathcal{H}}_0^{\text{SOI}}$ contribution outlined in orange and the full $\hat{\mathcal{H}}_0^{\text{SOI}}+\hat{\mathcal{H}}_1^{\text{SOI}}$ outlined in blue at the energy $E_0 = 1.65$ eV. (a2-3) Same as (a1) but at a fixed $k_x = 0.1~\mu\text{m}^{-1}$ and $k_y = 0.1~\mu\text{m}^{-1}$ respectively. (a4) Energy shift $\Sigma(\boldsymbol{k})$ dependence on $\vec{\boldsymbol{k}_\parallel}$. (a5) Optical birefringence $\alpha({\boldsymbol{k}})$ dependence on $\vec{\boldsymbol{k}_\parallel}$. (a6) Real part of the TE-TM splitting $\beta(\boldsymbol{k})(k_x+ik_y)^2$ dependence on $\vec{\boldsymbol{k}_\parallel}$. (a7) Imaginary part of the TE-TM splitting $\beta(\boldsymbol{k})(k_x+ik_y)^2$ dependence on $\vec{k_\parallel}$. (a8) Effective magnetic field texture $\mathbf G(\boldsymbol{k})$, with arrows representing $\mathbf G_\parallel(\boldsymbol{k})$ and the background color representing $G_z(\boldsymbol{k})=\delta_z(\boldsymbol{k})$. (b1-8) Illustrations of the same parameters as (a1-8) but using the dipole orientation of Model II. (c1-8) Illustrations of the same parameters as (a1-8) but using the dipole orientation of Model III and an energy of $E_0 = 1.6$ eV, with diabolical points indicated by yellow circles. (d1-8) Illustrations of the same parameters as (a1-8) but using the dipole orientation of Model IV and an energy of $E_0 = 1.8$ eV, with diabolical points indicated by yellow circles.} 
    \label{Fig3_comparing}
\end{figure*}
Figs.~\ref{fig2_Alldispersion}a2-a8 present the dispersions from different perspectives for Model I (i.e., ${\epsilon}_1 = {\epsilon}_2$, $|\vec{\boldsymbol{\mu}}_1| = |\vec{\boldsymbol{\mu}}_2|$, and $\vec{\boldsymbol{\mu}}_1 \perp \vec{\boldsymbol{\mu}}_2$) as illustrated in Fig. \ref{fig2_Alldispersion}a1. Such a setup represents a trivial bulk isotropic material, which enables no nontrivial spin-momentum locking. In this setup, the generalized Zeeman splitting term, $\delta_z(\boldsymbol{k}) = 0$, vanishes due to the absence of out-of-plane transition dipoles, and the Rashba-like displacement (illustrated in Fig.~\ref{Fig1_schemband}b) of the spin-resolved band minima is also absent as a result. In addition to this, the TE-TM splitting  vanishes with $\beta ({\boldsymbol{k}}) = 0$ since $\sum_d\left( \vec{\boldsymbol{\mu}}^d_\parallel \cdot \vec{\boldsymbol{e}}_- \right)^2 = 0$, as illustrated in Fig.~\ref{Fig3_comparing}a6-a7. Meanwhile, the parameters $\Sigma(\boldsymbol{k})$ and $\alpha(\boldsymbol{k})$ are isotropic (see Fig.~\ref{Fig3_comparing}a4-a5), depending only on $|\boldsymbol{k}|$. Specifically, the higher-order spin-independent self-energy $\Sigma_1(\boldsymbol{k}) \propto - |\boldsymbol{k}|^2$, thereby increasing the effective photonic mass (as expected in a dielectric environment). The generalized optical birefringence term $\alpha(\boldsymbol{k})$ hybridizes the two photonic polarizations and generates two paraboloids centered at $\boldsymbol{k}=0$, as shown in Fig. \ref{fig2_Alldispersion}a2. As a result, these two paraboloids are not spin-polarized within our native RHP/LHP basis (characterized via $S_3$, not shown here) and do not exhibit spin-momentum locking. However, the outer and inner paraboloids assume different linear spin-polarizations (characterized via $S_1$) as clearly seen in Fig.~\ref{fig2_Alldispersion}a3. The splitting vanishes at $|\boldsymbol{k}|=0$, and the overall impact of this splitting is marginal for the relevant range of photonic dispersion presented in Fig.~\ref{fig2_Alldispersion}a5-a8. Consequently, the effective magnetic field reduces to $\vec{\boldsymbol{G}}(\boldsymbol{k}) = (\alpha(\boldsymbol{k}), 0, 0)$, which is displayed in Fig.~\ref{Fig3_comparing}a8. These results also show that $\hat{\mathcal{H}}_\mathrm{GRD}$ can qualitatively and semi-quantitatively reproduce the results obtained from $\hat{H}_{\vec{\boldsymbol{\mu}}\cdot\vec{\boldsymbol{E}}}$. In Fig. \ref{Fig3_comparing}a1-3, we compare the dispersions of $\hat{\mathcal{H}}_\mathrm{GRD}$ and $\hat{\mathcal{H}}'_\mathrm{GRD} = \omega_k \mathds{1} + \hat{\mathcal{H}}_0^{\text{SOI}}$ (outlined in orange) for this model. Importantly, the higher-order corrections lift the degeneracy of the dispersion, and $n=0$ terms add only a uniform shift to the energies across $k$-space. Since the degeneracy of the $\hat{\mathcal{H}}'_\mathrm{GRD}$ dispersion is not lifted, a distinctive Stokes parameter structure is not observed. Here, $\Sigma(\boldsymbol{k})$ is symmetric and has a maximum value of $\Sigma_0$ at $\boldsymbol{k}=0$, and is displayed in Fig. \ref{Fig3_comparing}a4. $\alpha(\boldsymbol{k})$ has a max value of $\alpha(0)=0$ at $\boldsymbol{k}=0$ with isotropic $k$-dependence, and $\beta(\boldsymbol{k})=0$, as illustrated in Figs. \ref{Fig3_comparing}a5-7. The magnetic field texture is shown in Fig. \ref{Fig3_comparing}a8. Since $\beta(\boldsymbol{k})=0$, the in-plane components of the magnetic field are $G_x(\boldsymbol{k})=\alpha(\boldsymbol{k})$ and $G_y(\boldsymbol{k})=0$; therefore, the magnetic field is aligned parallel to the $x$-direction at all $k$-points. Since $\mu_z=0$ in this dipole configuration, $G_z(\boldsymbol{k})=0$, as shown in the color background of Fig. \ref{Fig3_comparing}a8.


Figs.~\ref{fig2_Alldispersion}b2-b8 present the photonic dispersion from different perspectives for Model II, which represents a uniaxial material (with only one transition dipole per unit cell), such as liquid crystals in the nematic phase~\cite{sedov2025engineering, sedov2024light}, rectangular nanoplatelets,~\cite{scott2017directed, gao2017cdse} and nanowires~\cite{lin2020giant, johnston2022polarization}, as illustrated in Fig. \ref{fig2_Alldispersion}b1. This setup also has no out-of-plane transition dipoles; therefore, $\delta_z(\boldsymbol{k})=0$, and there is no in-plane displacement of the band minima. The spin-independent self-energy $\Sigma(\boldsymbol{k})$ increases the effective photonic mass, similar to the previous scenario, but in an anisotropic fashion, as expected for such a uniaxial material. The anisotropic nature of $\Sigma(\boldsymbol{k})$ is shown in Fig. \ref{Fig3_comparing}b4. In this setup, both the generalized TE-TM splitting and the optical birefringence are non-zero, which leads to the hybridization of the photonic polarizations. Since the TE-TM splitting term $\beta(\boldsymbol{k})$ is accompanied by the factor $(k_x \pm i k_y)^2 = |\boldsymbol{k}|^2 e^{\pm 2i\phi(\boldsymbol{k})}$ for $\phi(\boldsymbol{k})=\tan^{-1}(k_y/k_x)$, its phase winds twice as $\boldsymbol{k}$ rotates once in momentum space, causing the character of the resulting bands to alternate across the four quadrants and enable spin-momentum locking. This can be observed in Fig.~\ref{fig2_Alldispersion}b2-b4, where the photonic spin-polarization is characterized via $S_2$. Note that, in contrast to Model I, the bands here remain split in the $\boldsymbol{k}\to 0$ limit, as illustrated in Fig.~\ref{fig2_Alldispersion}b5-b8, because the $1/|\boldsymbol{k}|^2$ dependence of $\beta_0(\boldsymbol{k})$ cancels the quadratic momentum dependence of $(k_x \pm i k_y)^2$. 

Figs. \ref{fig2_Alldispersion}b3-8 compare the dispersions of $\hat{H}_{\vec{\boldsymbol{\mu}}\cdot\vec{\boldsymbol{E}}}$ and $\hat{\mathcal{H}}_\mathrm{GRD}$ at the fixed energy of $E_0=1.65$ eV for Model II. Once again, $\hat{\mathcal{H}}_\mathrm{GRD}$ reproduces the low-energy dispersion bands for a large $\boldsymbol{k}$ window and also produces the same $S_2$ pattern as in the full light-matter Hamiltonian. Figs. \ref{Fig3_comparing}b1-3 compare the dispersions of $\hat{\mathcal{H}}_\mathrm{GRD}$ and $\hat{\mathcal{H}}'_\mathrm{GRD}$ (in blue boxes) for Model II. Here, the main contributor to the energy dispersion and its Stokes parameter polarization structure comes from the leading-order SOI. Since $|\beta_0(\boldsymbol{k})(k_x+ik_y)^2|$ is greater than both $\alpha(\boldsymbol{k})$ and $|\beta_1(k_x+ik_y)^2|$, the dispersion near $\boldsymbol{k}=0$ has negligible contribution from higher-order SOI. The reciprocal-space forms of $\alpha(\boldsymbol{k})$ and $\beta(\boldsymbol{k})$ in this model are shown in Fig. \ref{Fig3_comparing}b5-7. Here, $\beta(\boldsymbol{k})$ is purely real, and $(k_x+ik_y)^2$ induces a phase sign switch at $|k_x|=|k_y|$. Fig. \ref{Fig3_comparing}b8 presents the in-plane (vectors) and out-of-plane (color background) magnetic field texture for Model II. There is no diabolical point in this model, indicating a non-zero magnetic field filling the reciprocal-space.

Figs. \ref{fig2_Alldispersion}c2-8 present the dispersions from different perspectives for Model III, which represents an anisotropic material (i.e., ${\epsilon}_1 \not= {\epsilon}_2$, $|\vec{\boldsymbol{\mu}}_1| = |\vec{\boldsymbol{\mu}}_2|$ and $\vec{\boldsymbol{\mu}}_1 \perp \vec{\boldsymbol{\mu}}_2$), as illustrated in Fig. \ref{fig2_Alldispersion}c1. This model describes materials such as phosphorene~\cite{novko2021ab, sharma2018highly}, orthorhombic perovskites~\cite{as2025recent}, and van der Waals oxyhalides~\cite{xu2026giant}. In this model material, $\hat{\mathcal{H}}_\mathrm{GRD}$ has two diabolical points corresponding to $\delta_z(\boldsymbol{k})=\alpha(\boldsymbol{k})+\beta(\boldsymbol{k})(k_x+ik_y)^2=0$, with coordinates given by
\begin{equation}
k_x^{\text{DB}}=0, \quad k_y^{\text{DB}} = \pm \frac{\pi}{L}\sqrt{\frac{2\Delta_1\Delta_2(\Delta_1-\Delta_2)}{\Delta_1\Delta_2(\Delta_1+\Delta_2)+\omega_0(\Delta_1^2-\Delta_2^2)}}\, ,
\end{equation}
for $\vec{k}_{\text{DB}} = k_x^{\text{DB}}\vec{\boldsymbol{e}}_x+k_y^{\text{DB}}\vec{\boldsymbol{e}}_y$ (see SI for details). Fig. \ref{fig2_Alldispersion}c2 shows the isometric view of the dispersion with variation of the Stokes parameter $S_2$ at different energies. Figs. \ref{fig2_Alldispersion}c3-4 compare the dispersions of $\hat{\mathcal{H}}_{\vec{\boldsymbol{\mu}}\cdot \vec{\boldsymbol{E}}}$ and $\hat{\mathcal{H}}_\mathrm{GRD}$, with the locations of $k_{\text{DB}}$ in the $\hat{\mathcal{H}}_\mathrm{GRD}$ dispersion in close agreement with those of the $\hat{\mathcal{H}}_{\vec{\boldsymbol{\mu}}\cdot \vec{\boldsymbol{E}}}$ dispersion. Additionally, the two dispersions show the same Stokes parameter $S_2$ pattern, indicating that $\hat{\mathcal{H}}_\mathrm{GRD}$ should reproduce the original $\hat{\mathcal{H}}_{\vec{\boldsymbol{\mu}}\cdot \vec{\boldsymbol{E}}}$ Stokes parameter dynamics. Figs. \ref{Fig3_comparing}c1-3 compare the dispersions of $\hat{\mathcal{H}}_\mathrm{GRD}$ and $\hat{\mathcal{H}}'_\mathrm{GRD}$. Here, the dominant contributions to $\hat{\mathcal{H}}_\mathrm{GRD}$ are higher-order terms ($n>0$). Fig. \ref{Fig3_comparing}c1 shows that both terms generate the same Stokes parameter pattern, but the dispersion and diabolical points ultimately originate from the higher-order terms. Notably, as $k\rightarrow 0$, the zeroth-order term becomes dominant as shown in Figs. \ref{Fig3_comparing}c2-3. However, for larger wavevectors ($|{\boldsymbol{k}}|\gtrapprox 4\,\mu\text{m}^{-1}$), the higher-order correction terms dominate, regenerating $\hat{\mathcal{H}}_{\vec{\boldsymbol{\mu}}\cdot \vec{\boldsymbol{E}}}$ and its diabolical points. Even though $|\vec{\boldsymbol{\mu}}_1|=|\vec{\boldsymbol{\mu}}_2|$, $\Sigma(\boldsymbol{k})$ is anisotropic in $k$-space since $\epsilon_1\neq\epsilon_2$, as shown in Fig. \ref{Fig3_comparing}c4. Since $\alpha(\boldsymbol{k})$ and $\operatorname{Re}[\beta(\boldsymbol{k})]$ depend only on $|\vec{\boldsymbol{\mu}}|$ and $|\boldsymbol{k}|^2$, both are isotropic functions of $|\boldsymbol{k}|$, shown in \ref{Fig3_comparing}c5-7. Fig. \ref{Fig3_comparing}c8 shows the in-plane magnetic field texture, which vanishes near the diabolical points (at $k_y^{\text DB} = \pm 7.21~\mu\text{m}^{-1}$ represented by yellow circles) which arise when $\vec{\boldsymbol{G}}(\boldsymbol{k})=0$. In this model, since $G_z(\boldsymbol{k})=0$, there is no out-of-plane magnetic field texture.

Finally, Figs. \ref{fig2_Alldispersion}d2-8 present the dispersions from different perspectives for Model IV (i.e., ${\epsilon}_1 = {\epsilon}_2$, $|\vec{\boldsymbol{\mu}}_1| = |\vec{\boldsymbol{\mu}}_2|$, and $\vec{\boldsymbol{\mu}}_1(x,y,z) = \vec{\boldsymbol{\mu}}_2(-x,y,z)$), as depicted in Fig. \ref{fig2_Alldispersion}d1. This setup represents a canted material, such as liquid crystals in different electrical voltages \cite{liang2024polariton}. 
Unlike the previous models, these transition dipoles do have a component along the $z$-direction; as a result, $\delta_z({\boldsymbol{k}})\neq 0$ when $k_x\neq0$, as illustrated in Fig. \ref{fig2_Alldispersion}d8 (see SI). This material structure produces two diabolical points at $\beta(\boldsymbol{k^{\text{DB}}})=\alpha(\boldsymbol{k^{\text{DB}}})=\delta_z(\boldsymbol{k^{\text{DB}}})=0$ whose coordinates in $k$-space are written as
\begin{equation}
k_x^{\text{DB}}=0, \quad k_y^{\text{DB}} = \pm \frac{\pi}{L}\sqrt{\frac{\sum_d \left(2(\mu_z^d)^2 - |\vec{\boldsymbol{\mu}}^d_\parallel|^2\right)}{\sum_d (\mu_z^d)^2}}\, .
\end{equation}
Fig. \ref{fig2_Alldispersion}d2 shows the isometric view of the dispersion of this model, consisting of two paraboloids whose vertices are separated along the $x$-direction (see SI). The $S_3$ Stokes parameter of these two paraboloids has opposite signs and is independent of the $k$-space quadrant. This dispersion closely resembles the electronic Rashba-Dresselhaus Hamiltonian in the presence of Zeeman splitting with broken time-reversal symmetry. Figs. \ref{fig2_Alldispersion}d3-4 compare the dispersions of $\hat{\mathcal{H}}_{\vec{\boldsymbol{\mu}}\cdot \vec{\boldsymbol{E}}}$ and $\hat{\mathcal{H}}_\mathrm{GRD}$ at a fixed energy cut of $E_0=1.8$ eV. We again observe visually identical dispersions and Stokes parameter structure in our model and $\hat{\mathcal{H}}_{\vec{\boldsymbol{\mu}}\cdot \vec{\boldsymbol{E}}}$. Figs. \ref{Fig3_comparing}d1-3 compare the dispersions of $\hat{\mathcal{H}}_\mathrm{GRD}$ and  $\hat{\mathcal{H}}'_{\text{GRD}}$ for Model IV. In this model, the zeroth-order terms in $\hat{\mathcal{H}}'_{\text{GRD}}$ do not lift the RHP and LHP degeneracy for any $\boldsymbol{k}$. Here, $\Sigma(\boldsymbol{k})$ is isotropic in $\phi(\boldsymbol{k})$ with a maximum at $|\boldsymbol{k}|=0$, as shown in Fig.~\ref{Fig3_comparing}d4. In this model, $\alpha(\boldsymbol{k})$ is also isotropic in $\phi(\boldsymbol{k})$ and vanishes when $|\boldsymbol{k}|=10.55~\mu\text{m}^{-1}$, as presented in Fig.~\ref{Fig3_comparing}d5. This indicates that the diabolical point condition is fulfilled when $k_y^{\text{DB}}=\pm 10.55~\mu\text{m}^{-1}$ (represented by yellow circles in Fig.~\ref{Fig3_comparing}d8). Since $\epsilon_1=\epsilon_2$ and $|\vec{\boldsymbol{\mu}}_\parallel^1|=|\vec{\boldsymbol{\mu}}_\parallel^2|$, the TE-TM splitting vanishes with $\beta ({\boldsymbol{k}}) = 0$ as $\sum_d\left( \vec{\boldsymbol{\mu}}^d_\parallel \cdot \vec{\boldsymbol{e}}_- \right)^2 = 0$, which is shown in Figs.~\ref{Fig3_comparing}d6-7. Consequently, $G_y(\boldsymbol{k}) = 0$ and $G_x(\boldsymbol{k}) = \alpha(\boldsymbol{k})$, so the in-plane magnetic field vector direction flips when $\alpha(\boldsymbol{k})$ switches signs, as displayed in  Fig.~\ref{Fig3_comparing}d8. 

{\bf Conclusion.} In this work, we develop a rigorous theoretical framework to provide microscopic insights on the origins of spin-orbit coupling of light. Our approach goes beyond the phenomenological Rashba–Dresselhaus model by introducing a {\it generalized} Rashba–Dresselhaus description that provides crucial material-design principles for precisely tuning the optical birefringence $\alpha(\boldsymbol{k})$, TE–TM splitting $\beta(\boldsymbol{k})$, and Zeeman splitting term $\delta_z(\boldsymbol{k})$, thereby enabling the selective engineering of distinct SOI regimes. We demonstrate that virtual interactions to material excitonic states produce photonic spin-orbit couplings which can be modified by tuning excitonic energies, and the magnitude and orientation of their transition dipole moments. The analytical expressions developed in this work map exciton energies and transition-dipole orientations directly onto $\alpha(\boldsymbol{k})$, $\beta(\boldsymbol{k})$, and $\delta_z(\boldsymbol{k})$. We use four representative material model systems and discuss how different regimes of photonic SOI are generated depending on the electronic structure of the material system. Furthermore, we demonstrate the predictive power of our generalized (optical) Rashba-Dresselhaus model $\hat{\mathcal{H}}_\mathrm{GRD}$ by comparing numerical results obtained directly from the dipole gauge Hamiltonian $\hat{H}_{\vec{\boldsymbol{\mu}} \cdot \vec{\boldsymbol{E}}}$. 

Overall, our work sheds light on the impact of material design on optical spin-orbit interactions and provides a framework to systematically investigate novel material systems which target a desired photonic SOI regime. Our specific results encompass broad classes of photonic SOI architectures, building a foundation for developing components for optical technologies.\\\\

    

\begin{acknowledgments}
This work was supported by the U.S. National Science Foundation (NSF) under Grant
No. CHE-2611431 and the Office of Naval Research under award N00014-25-1-2079. This work used TAMU FASTER at Texas A\&M University through allocation  PHY230021 from the Advanced Cyberinfrastructure Coordination Ecosystem: Services \& Support (ACCESS) program, which is supported by National Science Foundation grants \#2138259, \#2138286, \#2138307, \#2137603, and \#2138296.  M.F. acknowledges support from the Texas Leadership Research Scholars (TLRS) scholarship. 
\end{acknowledgments}

\section*{Competing Interests}
The authors declare no competing interests.

}
\bibliography{bib.bib}
\bibliographystyle{naturemag}

\end{document}